\documentclass[12pt, preprint]{aastex}
\usepackage{graphicx}
        
\begin{document}

\title{On the formation of hot DQ white dwarfs}

\author{L. G. Althaus$^{1,2}$}
\author{E. Garc\'{\i}a--Berro$^{3,4}$}
\author{A. H. C\'orsico$^{1,2}$}
\author{M. M. Miller Bertolami$^{1,5}$}
\author{A. D. Romero$^{1,5}$}
\affil{$^1$Facultad de Ciencias Astron\'omicas y Geof\'{\i}sicas,
       Universidad  Nacional de  La Plata,  
       Paseo del  Bosque s/n,
       (1900) La Plata, 
       Argentina\\
       $^2$Member of the Carrera del Investigador Cient\'{\i}fico y 
       Tecnol\'ogico, CONICET (IALP), Argentina\\
       $^3$Departament de F\'\i sica Aplicada, 
       Escola Polit\`ecnica Superior de Castelldefels,
       Universitat Polit\`ecnica de Catalunya,  
       Av. del Canal Ol\'\i mpic, s/n,  
       08860 Castelldefels,  
       Spain\\
       $^4$Institut d'Estudis Espacials de Catalunya, 
       c/ Gran Capit\`{a} 2--4, 
       08034 Barcelona, 
       Spain\\
       $^5$ Fellow of CONICET}

\email{althaus@fcaglp.unlp.edu.ar}

\begin{abstract}
We present the first full evolutionary calculations aimed at exploring
the origin  of hot DQ  white dwarfs.  These  calculations consistently
cover the whole evolution from the born-again stage to the white dwarf
cooling  track.   Our  calculations  provide  strong  support  to  the
diffusive/convective-mixing  picture  for  the  formation of  hot  DQs.  
We find that the hot DQ stage is a
short-lived stage  and that the range of  effective temperatures where
hot DQ  stars are found  can be accounted  for by different  masses of
residual helium and/or different initial stellar masses.  In the frame
of this scenario, a  correlation between the effective temperature and
the  surface carbon  abundance in  DQs  should be  expected, with  the
largest  carbon abundances  expected  in the  hottest  DQs.  From  our
calculations, we suggest that most of  the hot DQs could be the cooler
descendants of  some PG1159  stars characterized by  He-rich envelopes
markedly  smaller  than those  predicted  by  the  standard theory  of
stellar evolution.   At least for  one hot DQ, the  high-gravity white
dwarf {\mbox  SDSS J142625.70+575218.4}, an  evolutionary link between
this star and the massive PG1159 star H1504+65 is plausible.
\end{abstract}



\section{Introduction}

White dwarf  stars constitute the  end-point of stellar  evolution for
the overwhelming  majority of stars.  In  fact, more than  97\% of all
stars in our  Galaxy are expected to end their  lives as white dwarfs.
Thus, the  present white  dwarf population contains  information about
the  history of  our Galaxy  and  has also  potential applications  as
reliable cosmic clocks  to infer the age of a  wide variety of stellar
populations --- see  Winget \& Kepler (2008) and  Fontaine \& Brassard
(2008) for recent reviews.

Traditionally,  white dwarfs  have been  classified into  two distinct
families  according to the  main chemical  constituents characterizing
their  surfaces:  those  with  a  H-dominated atmosphere  ---  the  DA
spectral type  --- which comprises  about 85\% of known  white dwarfs,
and  those with  a He-rich  surface composition  --- the  non-DA white
dwarfs --- which represent the  rest of the population. It is accepted
that most  non-DA white  dwarfs are the  direct descendants  of PG1159
stars ---  see Unglaub \& Bues  (2000) and Althaus et  al.  (2005) and
references therein ---  which are hot stars with  H-deficient and He-,
C- and O-rich surface layers  (Werner \& Herwig 2006). In fact, PG1159
stars constitute a transition  stage between the post-asymptotic giant
branch  (AGB)  stars and  most  of  the  H-deficient white  dwarfs.  A
significant fraction  of PG1159  stars are thought  to be formed  as a
result of  a born-again  episode, that is,  a very late  thermal pulse
(VLTP) experienced by a hot white dwarf during its early cooling phase
(Iben et  al. 1983; Herwig et.   al.  1999).  During the  VLTP, the He
flash-driven convection  zone reaches the H-rich envelope  of the star
and, consequently, most of the hydrogen is violently burned (Herwig et
al. 1999; Miller  Bertolami et al. 2006).  The star  is then forced to
evolve rapidly  back to the AGB and  finally as the central  star of a
planetary   nebula  at   high  effective   temperatures.   Ultimately,
gravitationally-induced  diffusion acting  during the  early  stage of
evolution leads to  the formation of a pure  He envelope, giving raise
to the DO and DB spectral  types --- see, for instance, Althaus et al.
(2005) and references therein.

The recent  discorvery of  a new lukewarm  population of  white dwarfs
with  C-rich atmospheres ---  known as  hot DQs  (Dufour et  al. 2007;
2008a)  --- has  sparked the  attention  of researchers  since it  has
raised the possibility of the  existence of a new evolutionary channel
of formation.  Dufour et al. (2008a) have reported that nine hot white
dwarfs identified in the Fourth  Data Release of the Sloan Digital Sky
Survey (SDSS)  are characterized  by atmospheres dominated  by carbon.
The existence  of these new  white dwarfs ---  all of them found  in a
narrow  effective temperature  strip (between  $\approx$ 18,000  K and
24,000 K) --- poses a challenge to the stellar evolution theory, which
cannot adequately explain  their origin. As proposed by  Dufour et al.
(2008a)  the  hot DQ  population  could be  related  to  the very  hot
($\approx 200,000$ K)  and massive ($0.83 \, M_{\sun}$)  member of the
PG1159  family,  the  enigmatic   star  H1504+65,  for  which  a  post
born-again   origin  is   not  discarded   (Althaus  et   al.   2009).
Interestingly enough, H1504+65  is the only known star  with no traces
of either  H or  He until the  discovery of  the hot DQ  white dwarfs.
Dufour et al.  (2008a) have outlined an evolutionary scenario in which
undetected traces of He remaining in the C- and O-rich outer layers of
H1504+65 would diffuse  upwards leading to a He-rich  white dwarf.  In
this  picture, a  C-rich atmosphere  should eventually  emerge  as the
result  of convective  mixing at  smaller effective  temperatures.  In
this  Letter, we  present the  first quantitative  assessment  of such
evolutionary scenario.  With the help of full evolutionary models that
consistently  cover the  evolution from  the born-again  stage  to the
cooling  track, we  present strong  theoretical evidence  supporting a
{\sl diffusive/convective mixing} picture like that proposed by Dufour
et al.  (2008a)  for the formation of hot DQs and  the existence of an
evolutionary link between these  stars and the PG1159 stars, including
H1504+65.


\section{Input physics and evolutionary sequences}

The calculations presented in this  work have been done using the {\tt
LPCODE} stellar  evolutionary code employed  in our previous  study of
the formation  of H-deficient post-AGB stars via  a born-again episode
(Althaus et al.  2005; Miller  Bertolami \& Althaus 2006). The code is
specifically designed to compute  the formation and evolution of white
dwarf  stars.  In  {\tt  LPCODE},  special emphasis  is  given to  the
treatment  of  the  changes  of  the  chemical  abundances,  including
diffusive  overshooting  and non-instantaneous  mixing,  which are  of
primary  importance  in the  calculation  of  the  thermal pulses  and
born-again stage that  lead to the formation of  PG1159 stars. We have
considered the following main physical ingredients.  Neutrino emission
rates  for pair, photo  and bremsstrahlung  processes were  taken from
Itoh et  al. (1996).  For  plasma processes we included  the treatment
presented  in  Haft et  al.   (1994).   The  OPAL radiative  opacities
(Iglesias \& Rogers 1996),  including C- and O-rich compositions, were
adopted. The conductive opacities are from Cassisi et al. (2007). This
prescription  covers the  entire regime  where electron  conduction is
relevant. In the low-density regime, we employed an updated version of
the  equation  of  state  of  Magni \&  Mazzitelli  (1979).   For  the
high-density regime, we use the  equation of state of Segretain et al.
(1994),  which accounts  for all  the important  contributions  in the
solid and  liquid phases.  We  have also considered  the gravitational
settling  and chemical  diffusion  of He,  C  and O  during the  whole
evolution. Our  treatment of time-dependent diffusion is  based on the
multicomponent  gas treatment presented  in Burgers  (1969). Diffusion
velocities  are evaluated  at each  evolutionary time  step.  Finally,
convection is treated within the formalism of the mixing length theory
as given by the ML2 parameterization (Tassoul et al. 1990).

The initial stellar models  needed to start our evolutionary sequences
correspond to realistic PG1159 stellar configurations derived from the
full evolution of their  progenitor stars (Miller Bertolami \& Althaus
2006).  We have  considered sequences with stellar masses  of 0.87 and
$0.585  \,  M_{\sun}$.  The  chemical  stratification  of our  initial
models consists of  a CO core, which is the result  of core He burning
in  prior stages,  surrounded by  a He-,  C- and  O-rich  envelope, in
agreement  to  what is  observed  in  PG1159  stars.  To  explore  the
possibility  that hot  DQ white  dwarfs can  be the  result  of mixing
events between the He-rich envelope and the underlying C-rich regions,
we have considered small masses  of residual He with fractional masses
in the  range $10^{-8}\le M_{\rm He}/M_{\rm WD}\le  2 \times 10^{-7}$.
The  sequence with  $0.87 \,  M_{\sun}$ was  specifically  computed to
explore the evolutionary connection  between H1504+65 and hot DQ white
dwarfs. For this  stellar mass, the He content  of $2 \times 10^{-7}\,
M_{\rm WD}$ corresponds to the maximum He content expected in H1504+65
if we assume a post-born-again  origin for this star (Miller Bertolami
\& Althaus 2006). For this sequence we assume an outer layer rich in C
and  O, with  trace  abundances  of He  reaching  deeper layers.   The
evolutionary   calculations   have    been   computed   from   $T_{\rm
eff}=100,000$ K  down to the domain of  effective temperatures typical
of hot DQs.

\begin{figure}
\begin{center}
\includegraphics[clip,width=250pt]{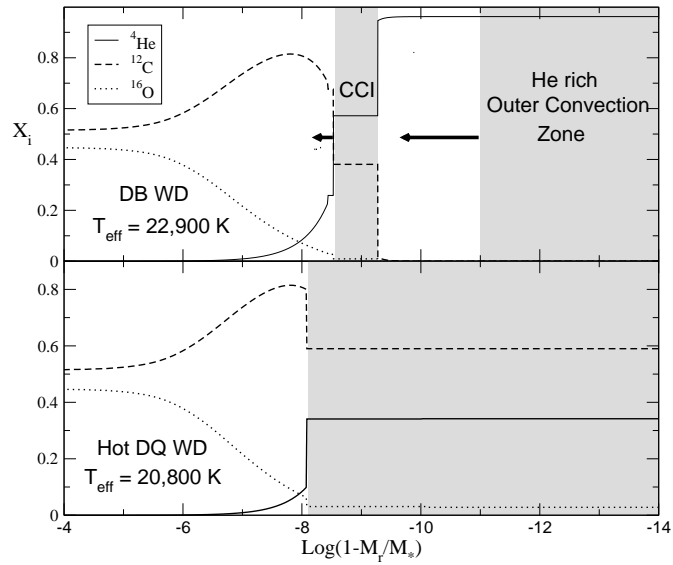}
\caption{Abundance distribution of $^{4}$He, $^{12}$C, and $^{16}$O as
         a  function  of  the  outer  mass fraction  at  two  selected
         effective temperatures  for the $0.87\,M_{\sun}$  white dwarf
         with  $M_{\rm He}= 10^{-8}\,M_{\rm  WD}$.  Gray  areas denote
         convectively   unstable  zones.   The   inward-growing  outer
         convection  zone  (upper panel)  merges  with the  underlying
         convective C-rich intershell  (CCI), leading to the formation
         of a  white dwarf  with a C  atmosphere ---  a hot DQ  --- at
         about $T_{\rm eff}= 20,800$ K (bottom panel).}
\label{quimi.eps}
\end{center}
\end{figure}


\section{Evolutionary results}

Fig \ref{quimi.eps}  illustrates the  main result of  our calculations
about the {\sl  diffusive/convective-mixing} picture for the formation
of hot DQs.   We show the results for the  $0.87 \, M_{\sun}$ sequence
with $M_{\rm He}= 10^{-8}\,M_{\rm  WD}$. As previously mentioned, this
stellar mass is similar to that  of H1504+65.  During the hot phase of
white  dwarf evolution,  the  chemical abundance  distribution of  the
envelope of  PG1159 stars is  strongly modified by  element diffusion.
Gravitational settling causes  He to float to the  surface and heavier
elements to sink.  In our models  when the white dwarf has cooled down
to  $T_{\rm eff}\approx  23,000$ K,  gravitationally-induced diffusion
has led to  the development of a pure He envelope  of nearly $7 \times
10^{-10}\,M_{\rm  WD}$, plus  an  extended tail  toward deeper  layers
(upper panel in  Fig \ref{quimi.eps}). Note that at  this stage of the
evolution, an inward-growing outer He convection zone is present. This
convective region is due to He recombination.

The large  values of  the radiative opacity  of the  C-enriched layers
below  the  pure He  envelope  produce  a  convective intershell  zone
stretching downward,  which increases the carbon  abundance there.  As
the  evolution proceeds,  the  base of  this  convective region  diggs
deeper into  C-rich layers, thus  further increasing the  abundance of
carbon in the intershell.  The existence of this convective intershell
is a key  aspect to understand the formation of hot  DQs.  It is worth
mentioning that the less massive  the He envelope, the larger the size
of the  convective intershell and the  larger the C  enrichment of the
intershell.   When $T_{\rm eff}=  21,000$ K,  the outer  He convection
zone reaches  the underlying convective  intershell, now substantially
enriched  in  C.   The   resulting  mixing  process  between  the  two
convective regions gives  rise to the formation of  a white dwarf with
outer layers rich in  He and C (and small traces of  O), that is a hot
DQ (bottom panel in Fig \ref{quimi.eps}).

Although the  evolution described  above is qualitatively  similar for
sequences  with different  He contents,  the effective  temperature at
which  mixing occurs  and the  surface chemistry  after  mixing depend
quite sensitively upon the exact value  of the residual He mass and on
the stellar  mass. In fact, the  surface C enrichment  of the emerging
hot  DQ depends on  the size  of the  convective intershell,  which is
larger  for the  case of  a smaller  He envelope.   These  results are
summarized in  Table 1 which, in  addition to the stellar  mass and He
content, lists the effective temperature and surface abundances of He,
C  and  O by  the  time  the outer  convection  zone  merges with  the
convective  intershell.   It is  clear  that  the  range of  effective
temperature where hot  DQs are found, as well  as the observed surface
chemistry, can be naturally accounted for by assuming different masses
of the residual He content with which white dwarfs enter their cooling
track  and/or different  white  dwarf masses.   Finally,  it is  worth
mentioning that the residual He  content required for this scenario to
work is  much larger than the value  of $10^{-15}\,M_{\sun}$ suggested
by Dufour et  al. (2008a), and consequently easier  to justify by prior
evolution.


\section{An evolutionary connection between H1504+65 and hot DQ white 
         dwarfs?}

Our  $0.87\,M_{\sun}$  sequence  is  of immediate  relevance  for  the
discussion of  an evolutionary link between the  massive H1504+65 star
and hot  DQs. Note from Table  1 that for  this sequence and for  a He
content between $M_{\rm He}\approx 10^{-8}$ and $10^{-7}\,M_{\rm WD}$,
the  formation of  a DQ  white dwarf  is expected  to occur  at nearly
$T_{\rm  eff}=20,000$  K, and  the  predicted  surface composition  is
expected to be  He and C in roughly equal fractions.   This is in very
good  agreement   with  the  observational  values   derived  for  the
high-gravity hot DQ white  dwarf {\mbox SDSS J142625.70+575218.4}, for
which Dufour et al.  (2008a) obtain an effective temperature of nearly
$T_{\rm eff}=20,000$  K and  a He- and  C-rich surface  composition of
log(C/He)=0.

\begin{table}
\caption{Main   characteristics  of   the  white   dwarf  evolutionary
         sequences by the time the  outer convection zone leads to the
         formation of C-rich outer layers.}  
\centering
\begin{tabular}{@{}ccllll}
\hline
\hline
$M_{\rm WD}/M_{\sun}$ & 
$M_{\rm He}/M_{\rm WD}$ & 
$T_{\rm eff}$ (K) & 
$^4$He & 
$^{12}$C & 
$^{16}$O\\
\hline 
0.870 & $10^{-8}$          & 20,700 & 0.36 & 0.58 & 0.03  \\
0.870 & $2 \times 10^{-7}$ & 17,300 & 0.63 & 0.31 & 0.02  \\
0.585 & $10^{-8}$          & 21,100 & 0.17 & 0.77 & 0.03  \\
0.585 & $2 \times 10^{-7}$ & 19,500 & 0.74 & 0.22 & 0.003 \\
\hline
\hline
\end{tabular}
\end{table}

\begin{figure}
\begin{center}
\includegraphics[clip,width=250pt]{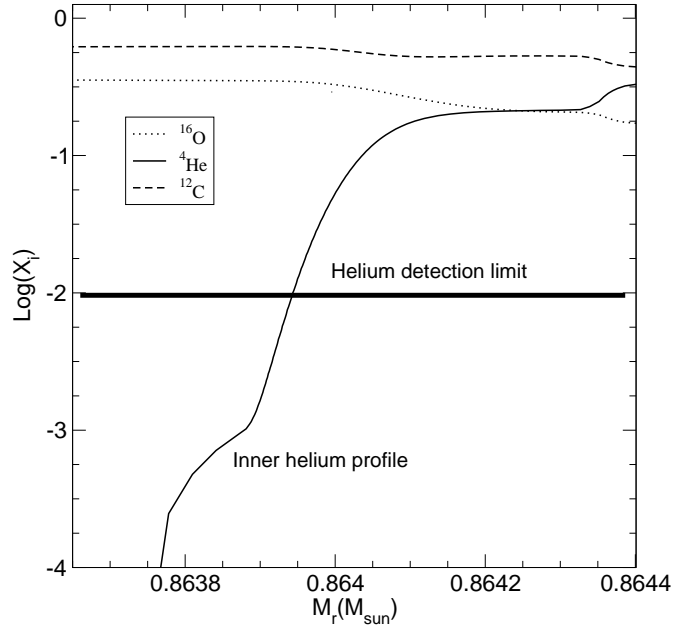}
\caption{The  internal abundance  distribution of  $^{4}$He, $^{12}$C,
         and  $^{16}$O for  a selected  model of  our  post born-again
         $0.87\,M_{\sun}$  sequence. The  mass coordinate  ranges from
         0.86365  to $0.8644\,M_{\sun}$.  Note  that a  non-negligible
         mass  of  residual He,  of  about  $\sim  2 \times  10^{-7}\,
         M_{\sun}$ is  expected below the layer having  a He abundance
         of 0.01, the He detection limit in H1504+65 (thick line).}
\label{perfil.eps}
\end{center}
\end{figure}

Is the amount of He required in this case consistent with the residual
He mass that could be expected  in a star like H1504+65? As mentioned,
H1504+65  is the  hottest and  most  massive known  PG1159 star.   Its
atmosphere is dominated by C and O plus traces of heavier elements ---
$X_{\rm  C}\simeq  0.48$, $X_{\rm  O}\simeq  0.48$, $X_{\rm  Ne}\simeq
0.02$,  $X_{\rm  Mg}\simeq 0.02$  (Werner  et  al.  2004).   Recently,
Althaus et  al. (2009) have explored  different evolutionary scenarios
for its formation and they do not discard a post born-again origin for
this  star as  a result  of non-stationary  mass-loss after  the VLTP.
From the  detailed simulations of  the born-again episode  for massive
remnants (Miller Bertolami  \& Althaus 2006) it is  found that even in
the case  in which most of  the He-rich envelope is  eroded during the
Sakurai stage, a non-negligible mass  of He remains in relatively deep
layers.   This  is  illustrated  in  Fig.   \ref{perfil.eps}  for  the
$0.87\,M_{\sun}$ post born-again sequence.  Note that the total amount
of residual He  below the layer having a He abundance  of 0.01 --- the
detection limit of He in H1504+65 (Werner et al.  2004) --- amounts to
$M_{\rm He}= 2  \times 10^{-7}\,M_{\sun}$, which can be  smaller if He
in the  surface layers of  H1504+65 exists with abundances  much lower
than the detection limit. This is  of the order of magnitude of the He
mass  required by  the  diffusive/convective-mixing picture  described
here  to  explain the  observational  characteristics  of {\mbox  SDSS
J142625.70+575218.4}. Thus, we have a consistent picture that provides
theoretical  support for an  evolutionary connection  between H1504+65
and this hot DQ white dwarf.


\section{Conclusions}

We have  presented full evolutionary calculations  that provide strong
support to a diffusive/convective-mixing picture like that proposed by
Dufour  et  al.   (2008a)  to  explain  the origin  of  hot  DQ  white
dwarfs. We have shown that mixing between the outer He convection zone
with the  underlying convective carbon intershell gives  rise to white
dwarf  structures  with  the  appropriate effective  temperatures  and
surface compositions  inferred from the observation of  hot DQs.  With
the help of detailed  born-again simulations for massive remnants, the
scenario  described  in  this  work  provides  the  first  theoretical
evidence   for  an   evolutionary  link   between  the   hydrogen  and
He-deficient PG1159 star H1504+65 with the high-gravity DQ white dwarf
{\mbox SDSS J142625.70+575218.4}, a connection that can be traced back
to strong mass-loss episodes after the VLTP.

For the intermediate-gravity DQ white dwarfs we find that the scenario
predicts  the  formation  of  C-dominated envelopes  for  residual  He
contents of the order of  $M_{\rm He}= 10^{-8}\,M_{\rm WD}$ or smaller
(see Table 1). As the residual He content is increased, it is expected
the  formation of  hot DQs  at lower  effective temperatures  and with
smaller C  abundances at their surfaces.  According  to this scenario,
we should thus expect  a correlation between the effective temperature
and the surface C abundance, with the largest C abundances expected in
the  hottest DQs.  This  trend appears  to  be in  agreement with  the
observational   inferences   (Dufour   et  al.    2008a).    Numerical
difficulties prevent  us from following  the further evolution  of our
sequences beyond the merger of  the two convection zones.  However, we
can make some estimations.  For  the evolutionary models shown in Fig.
\ref{quimi.eps}  we  note that  the  convection  zone  just after  the
formation of  the hot DQ  reaches as deep as  $10^{-8}\,M_{\sun}$.  At
the  base of  the convective  zone the  diffusion time  scale  is much
shorter (ranging from $\tau_{\rm dif} \approx 10^5$ to $10^6$ yr) than
the  cooling time scale  ($\tau_{\rm cool}=  T_{\rm c}/\dot  T_{\rm c}
\approx 10^8 $ yr). Hence, we expect that the hot DQ stage is indeed a
short-lived stage in  the evolution of a white  dwarf.  This being the
case, the range  of effective temperature where the  hot DQs are found
could be accounted for by assuming different masses of the residual He
with  which white  dwarfs  enter the  cooling  track and/or  different
stellar  masses.   However, it  cannot  be  discarded that  additional
mixing  episodes  could  develop  with  further  cooling.   Additional
calculations would be needed to test this possibility.

Finally,  we note  that the  origin of  intermediate-gravity  DQ white
dwarfs  is unclear.   As  suggested  by Dufour  et  al.  (2008a),  the
presence of these  hot DQs could be indicating the  existence of a new
evolutionary scenario for  the formation of white dwarfs.   In view of
the  evolutionary  connection between  H1504+65  and the  high-gravity
{\mbox  SDSS J142625.70+575218.4}, we  suggest that  the hot  DQ white
dwarf population of intermediate-gravity could also be the descendants
of some  PG1159 stars.  For  this to be  possible in the frame  of the
evolutionary  picture  described here,  some  PG1159  stars should  be
characterized  by  He  contents   orders  of  magnitude  smaller  than
predicted  by   the  standard  theory  of   stellar  evolution.   This
possibility  is not discarded  by the  recent theoretical  findings of
Althaus et  al.  (2008),  which suggest that  a thin  He-rich envelope
appears to be needed to solve the longstanding discrepancy between the
observed  (Costa \&  Kepler 2008)  and theoretical  (C\'orsico  et al.
2008) rates  of period  change of the  pulsating star  PG1159-035, the
prototype of the  PG1159 stars. This being the  true course of events,
we should  face the  problem of explaining  the coexistence  of PG1159
stars characterized by markedly  different He contents. In this sense,
the presence of a companion star  in a PG1159 star (Nagel et al. 2006)
may  be indicating  another channel  of formation  of these  stars. In
closing, it  is worth noting  that at least  for one hot DQ,  a strong
magnetic field has been detected  (Dufour et al.  2008b). As discussed
by these authors, the presence of this magnetic field might affect the
convection zones  significantly, thus altering the  predictions of the
diffusive/mixing scenario described here.

\begin{acknowledgements} 
Part of this  work was supported by the  MEC grant AYA05-08013-C03-01,
by the  European Union FEDER funds,  by the AGAUR,  by AGENCIA through
the Programa  de Modernizaci\'on Tecnol\'ogica BID  1728/OC-AR, and by
PIP 6521 grant from CONICET
\end{acknowledgements}

\end{document}